\documentstyle[aps]{revtex}
\begin{document}
\title{Simultaneous Brownian Motion of N Particles in a Temperature Gradient}

\author{J. M. Rub\'{\i}$^{a}$ and P. Mazur$^{b}$}
\address{$^{a}$Departament de F\'{\i}sica Fonamental, Facultat de F\'{\i}sica,\\ 
Universitat de Barcelona, Diagonal 647
08028 Barcelona, Spain\\
$^{b}$Instituu-Lorentz, University of Leiden, P.O. Box 9506, 2300 RA Leiden, The 
Netherlands.\\
}

\maketitle

\begin{abstract}
A system of N Brownian particles suspended in a nonuniform heat bath is treated 
as a thermodynamic system whith internal degrees of freedom, in this case their 
velocities and coordinates. Applying the scheme of non-equilibrium 
thermodynamics, one then easily obtains the Fokker-Planck equation for 
simultaneous Brownian motion of N particles in a temperature gradient. This 
equation accounts for couplings in the motion as a result of hydrodynamic 
interactions between particles. 
\end{abstract}


\section{Introduction}

In a former paper \cite{kn:agusti}, we reconsidered within the context of 
nonequilibrium thermodynamics for systems with internal degrees of freedom 
\cite{kn:groot}, the problem of Brownian motion in a temperature gradient and 
obtained in a simple way, for this case,the Fokker-Planck equation for one 
Brownian particle. This equation had been originally derived in 1968 by Zubarev 
and Bashkirov \cite{kn:zubarev} using the more elaborate statistical mechanical 
methods developed by Zubarev for treating nonequilibrium systems.

More recently Shea and Oppenheim \cite{kn:shea} applied statistical mechanical 
methods based on the use of projection operators to the study of Brownian motion 
of one particle in a nonequilibrium bath. They obtain a Fokker-Planck equation 
which, in the presence of a temperature gradient, agrees with the one derived by 
Zubarev and Bashkirov \cite{kn:zubarev}, and thus also with the one obtained 
from the nonequilibrium thermodynamic scheme for systems with internal degrees 
of freedom. The concept of internal degrees of freedom had been introduced 
already in 1953 into the formalism of nonequilibrium thermodynamics by Prigogine 
and Mazur \cite{kn:prigogine}, who pointed out that in this way 
Fokker-Planck-like kinetic equations could be derived.

In this paper we generalize our previous result to the motion of $\it{many}$ 
spherical Brownian particles suspended in a heat bath in the presence of a 
temperature gradient. In $\S$2 the system of N Brownian particles suspended in a 
nonuniform heat bath is considered as a system with internal degrees of freedom, 
consisting of the positions and the velocities of those particles. As a 
consequence, the probability density function in the phase space of the N 
Brownian particles plays the role of a thermodynamic density variable of a many 
component system, for which we give the Gibbs equation, containing the Massieu 
function conjugated to that density variable. Using Gibbs entropy postulate we 
are able to express this Massieu function in terms of the N Brownian particle 
distribution function. In $\S$3 we discuss the form of this distribution 
function in the state of local equilibrium, in which each particle has a 
velocity distribution in equilibrium with the temperature at its position.

We are then able not only to state in $\S$4 the conservation laws obeyed by the 
energy density and the densities of the various components with different  
internal degrees of freedom, but also to calculate the entropy production 
accompanying non uniformities, in ordinary coordinate space, as well as in the 
phase space of the internal degrees of freedom.

In $\S$5  we follow the principles  of nonequilibrium thermodynamics to 
formulate the linear phenomenological laws for the dissipative fluxes occurring 
in the entropy source strength. These phenomenological laws contain couplings 
between fluxes pertaining to different particles, which if realized find their 
origin in hydrodynamic interactions. Substitution of the phenomenological laws 
into the conservation equations gives rise on the one hand to the desired 
Fokker-Planck equation for N particles, and on the other to a coupled 
differential equation for the temperature distribution of the heat bath.

In $\S$6 we discuss the result obtained, also in comparison to - and in the 
light of - the merits of a more fundamental statistical mechanical approach. 
Moreover we address the problem why and when the inclusion of internal degrees 
of freedom in the formalism of nonequilibrium thermodynamics gives rise to valid 
and acceptable results.  

\section{The Brownian gas as a thermodynamic system with internal degrees of 
freedom}

Consider a gas of N Brownian particles of mass m suspended in a heat bath at 
rest with constant mass density $\rho_{H}$ and fixed total volume V. A 
nonuniform temperature field T({\bf r}) exists in the system (Brownian gas and 
heat bath) at position {\bf r}.

The Brownian particles are assumed to be essentially hard spheres and to have no 
further direct interactions. Their center of mass positions in the heat bath are 
${\bf R}_{\it{i}}$, $\it{i}$\,=\,1,2,..,N, their velocities ${\bf U}_{i}$. The 
probability density for the Brownian particles to be at time $\it{t}$ in a state 
defined by a point ${\bf \Gamma}\,=\,({\bf R}^{N},{\bf U}^{N})\,=\,({\bf 
R}_{1},{\bf R}_{2},..,{\bf R}_{N};{\bf U}_{1},{\bf U}_{2},..,{\bf U}_{N})$ in 
the 6N dimensional velocity-coordinate space will be denoted by $P^{N}({\bf 
\Gamma},t)$ and satisfies the normalization condition

\begin{equation}\label{eq:b1}
\int\,P^{N}({\bf \Gamma},t)d{\bf \Gamma}\,=\,1.
\end{equation}

\noindent The quantity $NmP^N$ then represents the mass density of Brownian 
particles having velocities ${\bf U}^N$ and positions ${\bf R}^N$.

The system itself, Brownian gas and heat bath, has mass density
 
\begin{equation}\label{eq:b2}
\rho({\bf r},t)\,=\,\rho_{H}\,+\,\rho_{B}({\bf 
r},t)\,\equiv\,\rho_{H}+\int\,d{\bf \Gamma}\sum_{i}m\delta({\bf R}_{i}-{\bf r}) 
P^{N}({\bf \Gamma},t)\,=\,\rho_{H}\,+\,\sum_{i}\langle m\delta({\bf R}_{i}-{\bf 
r})\rangle .
\end{equation}                      

\noindent Here $\rho_{B}({\bf r},t)$ is the mass density of the Brownian gas; 
the bracket $\langle ..\rangle$ denotes an average over the distrubution 
function $P^{N}({\bf \Gamma},t)$.

The coordinates ${\bf R}_{\it{i}}$ and velocities ${\bf U}_{\it{i}}$ may be 
considered as 6N intermal degrees of freedom of the system, in the sense of 
nonequilibrium thermodynamics \cite{kn:prigogine}. Taking this point of view, 
which implies that the system can be looked at as a system of many components 
with mass densities $NmP^N({\bf \Gamma},t)$, the Gibbs equation for the total 
entropy S of the system may be written as

\begin{equation}\label{eq:b3}
\delta S\,=\,\int\,d{\bf r}\frac{1}{T({\bf r})}\delta\rho e\,-\,Nm\int\,d{\bf 
\Gamma}\Theta({\bf \Gamma},t)\delta P^{N}({\bf \Gamma},t),
\end{equation}

\noindent where $e({\bf r},t)$ is the energy per unit of mass and $\Theta$ the 
Massieu function conjugate to the density $NmP^N$. In writing down equation 
(\ref{eq:b3}) account has been taken of the fact that $\rho_H$ is constant.

Before carrying out the program of nonequilibrium thermodynamics (which is to 
calculate in the first place the entropy production occurring in the system as a 
consequence of the nonequilibrium distribution  of temperature and internal 
degrees of freedom) and thus to derive the Fokker-Planck equation for N Brownian 
particles in the presence of a temperature gradient, we need to establish an 
expression for the Massieu function $\Theta({\bf \Gamma})$, relating it to the 
(thermodynamic) variables $P^{N}({\bf \Gamma},t)$. For this purpose we make use 
of Gibbs entropy postulate and write the entropy $S$ whith respect to its local 
equilibrium value in terms of a distribution over possible internal states

\begin{equation}\label{eq:b4}
S\,=\,k\int\,d{\bf \Gamma} P^{N}({\bf \Gamma},t)\ln\frac{P^{N}({\bf 
\Gamma},t)}{P^{N}_{l.e.}({\bf \Gamma})}\,+\,S^{l.e.}.
\end{equation}

\noindent Here $P^{N}_{l.e.}({\bf \Gamma})$ and $S^{l.e.}$ are the distribution 
function and the entropy at local equilibrium;  k is Boltzmann's constant. By 
local equilibrium we mean  that each Brownian particle has at its position a 
distribution over internal (velocity ) states in equilibrium with respect to the 
temperature at that point. The entropy $S^{l.e.}$, which  depends only on the 
temperature distribution and the density distribution of Brownian particles in 
the heat bath, obeys the standard Gibbs equation

\begin{equation}\label{eq:b5}
\delta S\,=\,\int\,d{\bf r}(\frac{1}{T}\delta\rho 
e\,-\,\frac{\mu_{B}}{T}\delta\rho_{B}),
\end{equation}

\noindent where $\mu_{B}({\bf r})$ is the chemical potential of the Brownian gas 
per unit of mass in internal (local) equilibrium at temperature $T({\bf r})$ and 
density $\rho_{B}$.
 
On  the other hand we have from expression (\ref{eq:b4}) for the differential of 
S 

\begin{equation}\label{eq:b6}
\delta S\,=\,-k\int\,d{\bf \Gamma} \ln\frac{P^{N}({\bf 
\Gamma},t)}{P^{N}_{l.e.}({\bf \Gamma})}\delta P^{N}({\bf \Gamma},t)\,+\,\delta 
S^{l.e.}.  
\end{equation}

\noindent Comparing  Eqs. (\ref{eq:b3}) and (\ref{eq:b6}) we find using also 
Eqs. (\ref{eq:b2}) and (\ref{eq:b5})

\begin{equation}\label{eq:b7}
\Theta({\bf \Gamma},t)\,=\,\frac{k}{Nm}ln
\frac{P^{N}({\bf \Gamma},t)}{P^{N}_{l.e.}({\bf \Gamma})}\,+\, \Theta^{l.e.}({\bf 
\Gamma}),
\end{equation}

\noindent with

\begin{equation}\label{eq:b8}
\Theta^{l.e.}({\bf \Gamma})\,=\,\frac{1}{N}\sum_{i}\frac{\mu_{B}({\bf 
R}_{i})}{T({\bf R}_{i})}
\end{equation}

Thus in local equilibrium the Massieu function $\Theta^{l.e.}({\bf \Gamma})$ is 
uniform in internal velocity space.

\section{The N particle distribution function at local equilibrium}

It remains to determine the local equilibrium distribution function $P^{N}({\bf 
\Gamma},t)$. We first note that in full equilibrium (in the absence of a 
temperature gradient) $P^{N}$ is given by the canonical ensemble distribution 
function

\begin{equation}\label{eq:c1}
P^{N}_{eq}\,=\,\frac{1}{N!}exp\{(F^{N}-H^{N})/kT\},
\end{equation}
 
\noindent with $T$ the equilibrium temperature of the heat bath, $H^{N}$ the 
energy of the N Brownian particles and $F^{N}$ the N particle free energy

\begin{equation}\label{eq:c2}
H^{N}\,=\,\sum_{i}\frac{1}{2}mU^{2}_{i},
\end{equation}

\begin{equation}\label{eq:c3}
F^{N}\,=\,F^{N}(T,V)\,=\,Nmf_{B}(T,V/Nm).
\end{equation}

\noindent Here $f_{B}$ is the free energy of the Brownian gas per unit of mass 
which (for N sufficiently large) is a function of temperature and density only.

In Eq.(\ref{eq:c2}) the direct interactions between Brownian particles (even 
their hard sphere interactions) are neglected, as the system is assumed to be 
sufficiently dilute.

We shall now take $P^{N}$ at local equilibrium to be of the form

\begin{equation}\label{eq:c4}
P^{N}_{l.e.}\,=\,\frac{1}{N!}exp\{(mf_{i}-H_{i})/kT({\bf R}_{i})\},
\end{equation}

\noindent with

\begin{equation}\label{eq:c5}
f_{i}\,=\,f_{B}(T({\bf R}_{i}),\rho_{e})\, ,\;\; \rho_{e}\,\equiv\,\frac{Nm}{V},
\end{equation}

\begin{equation}\label{eq:c6}
H_{i}\,=\,\frac{1}{2}mU^{2}_{i},
\end{equation}

\noindent where $T({\bf r})$ is a given non uniform temperature field and 
$\rho_{e}$ the uniform equilibrium particle density.

The free energy $f_{i}$, which is determined by inserting (12) into (1) and 
using Stirling's approximation, \mbox{$\ln N!\,=\,N\ln N\,-\,N$}, has the 
classical form

$$f_{i}\,=\,\frac{kT({\bf 
R}_{i})}{m}[\ln\frac{\rho_{e}}{m}\,-\,\frac{3}{2}\ln\frac{2\pi kT({\bf 
R}_{i})}{m}\,-\,1]=$$
\begin{equation}\label{eq:c7}
=\,\mu_{B}({\bf R}_{i})\,-\,p_{B}({\bf R}_{i})\rho_{e}^{-1}.
\end{equation}

\noindent Here $\mu_{B}({\bf r})$ is the explicit form of the previously 
introduced chemical potential per unit of mass of the ideal Brownian gas and 
$p_{B}({\bf r})\,=\,kT({\bf r})\rho_{e}/m$ its - perfect gas - pressure.

Thus $P^{N}_{l.e.}$ can be written in the form 

\begin{equation}\label{eq:c8}
P^{N}_{l.e.}\,=\,\prod_{i}N^{-1}exp\{\frac{m[\mu_{B}({\bf 
R}_{i})-\frac{1}{2}{\bf U}_{i}^{2}]}{kT({\bf R}_{i})}\}.
\end{equation}

\noindent If expression (\ref{eq:c8}) is inserted into Eq.(\ref{eq:b7}) we 
obtain for the Massieu function $\Theta({\bf \Gamma},t)$ the alternative form

\begin{equation}\label{eq:c9}
\Theta({\bf \Gamma},t)\,=\,\frac{k}{Nm}\ln P^{N}({\bf 
\Gamma},t)\,+\,\frac{1}{N}\sum_{i}\frac{m{\bf U}_{i}^{2}}{2T({\bf 
R}_{i})}\,+\,\frac{k}{m}\ln N.
\end{equation}

\noindent We are now in a position to carry out the nonequilibrium 
thermodynamics analysis which leads to the derivation of the N particle 
Fokker-Planck equation. In the next section we shall establish the basic 
equations necessary for this derivation.

\section{Conservation laws and entropy production.}

Let us calculate the rate of change of entropy by differentiating 
Eq.(\ref{eq:b3}) with respect to time

\begin{equation}\label{eq:d1}
\frac{dS}{dt}\,=\,\int\,d{\bf r}\frac{1}{T({\bf r})}\frac{\partial\rho 
e}{\partial t}\,-\,Nm\int\,d{\bf \Gamma}\Theta({\bf \Gamma},t)\frac{\partial 
P^{N}({\bf \Gamma},t)}{\partial t}.
\end{equation}

\noindent To proceed with the analysis we need the conservation laws obeyed by 
the energy density $\rho e$ and the distribution function $P^{N}({\bf 
\Gamma},t)$.

We shall first consider the latter: in ${\bf \Gamma}$ space $P^{N}$ obeys the 
continuity equation

\begin{equation}\label{eq:d2}
\frac{\partial P^{N}}{\partial t}\,+\,\sum_{i}{\bf U}_{i}\cdot\frac{\partial 
P^{N}}{\partial {\bf R}_{i}}\,=\,\sum_{i}\frac{\partial }{\partial {\bf 
U}_{i}}\cdot {\bf J}_{{\bf U}_i}\; ,
\end{equation}

\noindent where the ${\bf J}_{{\bf U}_{i}}$ are components of  fluxes in 3N - 
dimensional velocity space, which account for the interaction of the Brownian 
particles with the heat bath. We have neglected all direct interactions 
$\it{between}$ these particles also in writing down Eq.(\ref{eq:d2}). This is a 
reasonable approximation since in our dilute system the particle dynamics are 
overwhelmingly determined by their dissipative interactions with the heat bath. 
It is indeed our aim to use nonequilibrium thermodynamics to find the laws 
obeyed by the fluxes ${\bf J}_{{\bf U}_{i}}$ and thus to derive the equation 
describing the resulting  N particle random motion problem (also when the heat 
bath is not uniform).

Next to Eq.(\ref{eq:d2}) we need the law of conservation of energy which can be 
formulated as follows

\begin{equation}\label{eq:d3}
\frac{\partial\rho e}{\partial t}\,=\,-\nabla\cdot {\bf J}_{q},
\end{equation}

\noindent where ${\bf J}_{q}$ represents a heat flux defined in the reference 
frame in which the heat bath is at rest.

Substituting in Eq. (\ref{eq:d1}) the conservation laws (\ref{eq:d2}) and 
(\ref{eq:d3}) we can write the rate of change of entropy as the sum of two terms

\begin{equation}\label{eq:d4}
\frac{dS}{dt}\,=\,\frac{d_{e}S}{dt}\,+\,\frac{d_{i}S}{dt}.
\end{equation}

\noindent The first term $\frac{d_{e}S}{dt}$ is the entropy supplied to the 
system, and is given by

\begin{equation}\label{eq:d5}
\frac{d_{e}S}{dt}\,=\,-\int\,d{\bf \Omega}\cdot\frac{{\bf J}_{q}}{T},
\end{equation}

\noindent while the second term $\frac{d_{i}S}{dt}$ represents the total entropy 
production and is equal to

\begin{equation}\label{eq:d6}
\frac{d_{i}S}{dt}\,=\,\int\,d{\bf \Gamma}\sum_{i}\{m\rho_{B}^{-1}({\bf 
R}_{i})P^N({\bf \Gamma}){\bf J}_{q,i}({\bf R}_{i})\cdot\nabla T^{-1}({\bf 
R}_{i})\,-\,{\bf J}_{{\bf U}_{i}}\cdot k\frac{\partial}{\partial {\bf 
U}_{i}}\ln\frac{P^{N}}{P^{N}_{l.e.}}\}.
\end{equation}

In Eq.(\ref{eq:d5}) $\Omega$ is the surface of the heat bath and $d{\bf \Omega}$ 
a vector of magnitude $d\Omega$ normal to the surface pointing outward. The 
vectors ${\bf J}_{q,i}$ in Eq.(\ref{eq:d6}) represent modified heat fluxes at 
the positions of the Brownian particles and are defined as 

\begin{equation}\label{eq:d7}
{\bf J}_{q,i}({\bf R}_{i},{\bf U}_{i})\,=\,J_{q}({\bf R}_{i})\,-\,\rho_{B}({\bf 
R}_{i})\frac{1}{2}{\bf U}_{i}^{2}{\bf U}_{i}.
\end{equation}

\noindent They contain contributions from the kinetic energy flows of the 
Brownian particles themselves.

In deriving Eq. (\ref{eq:d5}) and (\ref{eq:d6}) use has been made of:
\begin{itemize}
\item[1-] Gauss' theorem in coordinate space and the fact that the Brownian 
particles are confined           to the heat bath,\\
\item[2-] the fact that upon partial integration in velocity space the fluxes 
${\bf J}_{{\bf U}_{i}}$ vanish as           the velocities $\pm\,{\bf U}_{i}$ 
tend to infinity; and\\
\item[3-] the form(s) (\ref{eq:b7}) and/or (\ref{eq:c9}) of the Massieu function 
$\Theta({\bf \Gamma})$.
\end{itemize}

The entropy production (23), which is a positive quantity in accordance with the 
second law of thermodynamics, is written as an integral over the N-Brownian 
particle phase space. We shall now demand, in agreement with the extension of 
nonequilibrium thermodynamics to systems with internal degrees of freedom 
\cite{kn:prigogine}, that not only the integral (\ref{eq:d6}) be positive 
definite, but also its integrand itself,

\begin{equation}\label{eq:d8}
\frac{d_{i}S}{dt}\,=\,\int\,\sigma_{\Gamma}d{\bf \Gamma},
\end{equation}

\begin{equation}\label{eq:d9}
\sigma_{\Gamma}\,=\,-\sum_{i}m\rho_{B}^{-1}({\bf R}_{i})P^{N}{\bf J}_{q,i}\cdot 
T_{i}^{-2}\nabla T_{i}\,-\,\sum_{i} {\bf J}_{{\bf U}_{i}}\cdot 
k\frac{\partial}{\partial {\bf U}_{i}}\ln (\frac{P^{N}}{P^{N}_{l.e.}})\; \geq 
\,0.
\end{equation}

\noindent We have introduced here the notation $T_{i}\,\equiv\,T({\bf R}_{i})$.
The quantity $\sigma_{\Gamma}$, the entropy source strength in $\Gamma$ space, 
is a sum of products of dissipative fluxes ${\bf J}_{q,i}$ and ${\bf J}_{{\bf 
U}_{i}}$ respectively, and their conjugate thermodynamic forces, the gradients 
of temperature in coordinate space and of the Massieu function $\Theta({\bf 
\Gamma})$, cf. Eq.(\ref{eq:b7}), in velocity space.

\section{Phenomenological relations and the Fokker-Planck equation for 
N-Brownian particles.} 

Following the scheme of nonequilibrium thermodynamics \cite{kn:groot} we shall 
now establish linear phenomenological relations between the fluxes and 
thermodynamic forces occurring in expression (\ref{eq:d9}) for $\Gamma$. Since 
the system is isotropic these relations are

\begin{equation}\label{eq:e1}
m\rho_{B}^{-1}P^{N}{\bf J}_{q,i}\,=\,-\sum_{j}L_{T_{i}T_{j}}T_{j}^{-2}\nabla 
T_{j}\,-\,k\sum_{j}L_{T_{i}{\bf U}_{j}}\frac{\partial}{\partial {\bf U}_{j}}\ln 
(\frac{P^{N}}{P^{N}_{l.e.}})
\end{equation}

\begin{equation}\label{eq:e2}
{\bf J}_{{\bf U}_{i}}\,=\,-\sum_{j}L_{{\bf U}_{i}T_{j}}T_{j}^{-2}\nabla 
T_{j}\,-\,k\sum_{j}L_{{\bf U}_{i}{\bf U}_{j}}\frac{\partial}{\partial {\bf 
U}_{j}}\ln (\frac{P^{N}}{P^{N}_{l.e.}})
\end{equation}

Assuming Onsager's principle of microscopic reversibility to hold for the system 
studied\footnote{Onsager's theorem is based on the property of microscopic 
reversibility which holds for fluctuations, in particular, $\it{around}$         
 $\it{equilibrium}$. But in the system studied we would have to consider 
correlations of particle velocity fluctuations at different positions and 
temperatures around a quasi-stationary nonequilibrium state, for which the 
theorem does not necessarily hold. Neverthless it seems legitimate to use the 
theorem when one takes into account that the velocity correlations at two 
different position i and j, which find their origin in hydrodynamic 
interactions, are sufficiently short ranged in space for temperatures to be 
taken equal as in equilibrium.}, the coefficients in Eqs. (\ref{eq:e1}) and 
(\ref{eq:e2}) satisfy the Onsager Casimir reciprocal relations.

\begin{equation}\label{eq:e3}
L_{T_{i}T_{j}}\,=\,L_{T_{j}T_{i}},
\end{equation}

\begin{equation}\label{eq:e4}
L_{{\bf U}_{i}T_{j}}\,=\,-L_{T_{j}{\bf U}_{i}},
\end{equation}           

\begin{equation}\label{eq:e5}
L_{{\bf U}_{i}{\bf U}_{j}}\,=\,L_{{\bf U}_{j}{\bf U}_{i}}.
\end{equation}

\noindent Defining the heat conductivity coefficients $\lambda_{ij}$, the 
thermal acceleration coefficients $\gamma_{ij}$ and the friction coefficients 
$\beta_{ij}$

\begin{equation}\label{eq:e6}
\lambda_{ij}\,\equiv\,\frac{L_{T_{i}T_{j}}\rho_{B}}{mP^{N}T_{j}^2},
\end{equation}
           
\begin{equation}\label{eq:e7}
\gamma_{ij}\,\equiv\,\frac{L_{{\bf U}_{i}T_{j}}}{P^{N}T_{j}},
\end{equation}

\begin{equation}\label{eq:e8}
\beta_{ij}\,\equiv\,\frac{mL_{{\bf U}_{i}{\bf U}_{j}}}{P^{N}T_{j}},
\end{equation}

\noindent relations (\ref{eq:e1}) and (\ref{eq:e2}) become taking into account 
also (\ref{eq:e4}), as well as the fact that in (\ref{eq:e7}) 
$T_{j}\,\simeq\,T_{i}$ in good approximation (see also note previous page)

\begin{equation}\label{eq:e9}
m\rho_{B}^{-1}P^{N}{\bf 
J}_{q,i}\,=\,-\sum_{j}m\lambda_{ij}\rho_{B}^{-1}P^{N}\nabla T_{j}\,+\,\sum_{j}
m\gamma_{ji}(P^{N}{\bf U}_{j}+\frac{kT_{j}}{m}\frac{\partial P^{N}}{\partial 
{\bf U}_{j}}),
\end{equation}

\begin{equation}\label{eq:e10}
{\bf J}_{{\bf U}_{i}}\,=\,-\sum_{j}m\gamma_{ij}P^{N}\nabla 
T_{j}/T_{j}\,-\,\sum_{j}m\beta_{ij}(P^{N}{\bf 
U}_{j}+\frac{kT_{j}}{m}\frac{\partial P^{N}}{\partial {\bf U}_{j}}).
\end{equation} 

Eqs. (\ref{eq:e9}) and (\ref{eq:e10}) (as did Eqs.(\ref{eq:e1}) and 
(\ref{eq:e2})) give expression to the fact that the dissipative fluxes at 
different (particle) positions can be coupled to each other: the coefficients 
$\beta_{ij}$ represent the well-known mutual friction coefficients of spherical 
bodies which result  (and are calculated) from their hydrodynamic interactions 
as they move  through the fluid they are suspended in. Likewise the coefficients 
$\gamma_{ij}$ and $\lambda_{ij}$, $i\,\neq\,j$ could arise from hydrodynamic 
interactions. We do not know whether such effects have previously been found. As 
far as the heat conductivity matrix is concerned its dominant contribution to 
$\lambda_{ii}$ is the heat conductivity of the heat bath itself, while all other 
contributions and matrix elements arise from particle motions.

We shall now first formulate, using Eq.(\ref{eq:e9}), the phenomenological law 
obeyed by the total heat current in the heat bath. For this purpose we multiply 
both members of (\ref{eq:e9}) by $\delta({\bf R}_{i}-{\bf r})$, sum over all 
particles i, and integrate over ${\bf \Gamma}$-space. We then obtain the 
following law

\begin{equation}\label{eq:e11}
{\bf J}'_{q}({\bf r})\,=\,-\int\,\lambda({\bf r},{\bf r}')\nabla T({\bf 
r}')d{\bf r}'\,+\,\int\,\gamma({\bf r},{\bf r}')\rho_{B}({\bf r}'){\bf 
V}_{B}({\bf r}')d{\bf r}'.
\end{equation}

\noindent Here the total heat flux ${\bf J}'_{q}$ at ${\bf r}$ is given by 
($\it{cf.}$ Eq.(\ref{eq:d7}))

\begin{equation}\label{eq:e12}
{\bf J}'_{q}({\bf r})\,=\,{\bf J}_{q}({\bf 
r})\,-\,\sum_{i}\langle\frac{1}{2}m{\bf U}_{i}^2{\bf U}_{i}\delta({\bf 
R}_{i}-{\bf r})\rangle,
\end{equation}

\noindent while ${\bf V}_{B}({\bf r}')$, is the mean particle velocity,

\begin{equation}\label{eq:e13}
{\bf V}_{B}({\bf r})\,=\,\rho_{B}^{-1}\langle\sum_{i}m{\bf U}_{i}\delta({\bf 
R}_{i}-{\bf r})\rangle.
\end{equation}

\noindent The heat conductivity kernels $\lambda({\bf r},{\bf r}')$ and 
$\gamma({\bf r},{\bf r}')$ are defined by the following averages

\begin{equation}\label{eq:e14}
\lambda({\bf r},{\bf 
r}')\,=\,m\rho_{B}^{-1}\langle\sum_{ij}\lambda_{ij}\delta({\bf R}_{i}-{\bf 
r})\delta({\bf R}_{j}-{\bf r}')\rangle,
\end{equation}

\begin{equation}\label{eq:e15}
\gamma({\bf r},{\bf r}')\rho_{B}({\bf r}'){\bf V}_{B}({\bf 
r}')\,=\,\langle\sum_{ij}\gamma_{ij}m{\bf U}_{i}\delta({\bf R}_{i}-{\bf 
r})\delta({\bf R}_{j}-{\bf r}')\rangle.
\end{equation}

These kernels, which for ${\bf r}\,\neq\,{\bf r}'$ result as stated above from 
hydrodynamic interactions, can be assumed to be sufficiently short ranged, so 
that from a truly macroscopic point of view Eq. (\ref{eq:e11}) has the 
approximate form

\begin{equation}\label{eq:e16}
{\bf J}'_{q}\,=\,-\lambda\nabla T\,+\,\gamma\rho_{B}{\bf V}_{B},
\end{equation}

\noindent with the macroscopic heat conductivity 

\begin{equation}\label{eq:e17}
\lambda\,=\,\int\,\lambda({\bf r},{\bf r}')d{\bf r}',
\end{equation}

\noindent and

\begin{equation}\label{eq:e18}
\gamma\,=\,\int\,\gamma({\bf r},{\bf r}')d{\bf r}'.
\end{equation}

\noindent The last quantity is related to the thermal diffusion coefficient  of 
the Brownian gas (for the exact relation of $\gamma$ to the thermal diffusion 
coefficient see eg. ref. \cite{kn:agusti}). 

Substitution of Eq.(\ref{eq:e16}) into the law of conservation of energy 
(\ref{eq:d3}) gives rise to a differential equation for the temperature which is 
coupled to the first few Brownian particle velocity  moments (cf. also Eq. 
(\ref{eq:e12}). These moments can be calculated from the N particle distribution 
function $P^{N}({\bf \Gamma},t)$, which in turn is the solution of the 
differential equation obtained by substitution of Eq. (\ref{eq:e10}) into Eq. 
(\ref{eq:d2})  
 
\begin{equation}\label{eq:e19}
\frac{\partial P^N}{\partial t}\,+\,\sum_{i}{\bf U}_{i}\cdot\frac{\partial 
P^N}{\partial {\bf R}_{i}}\,=\,\sum_{ij}\frac{\partial}{\partial {\bf 
U}_{i}}\cdot [\beta_{ij}(P^{N}{\bf U}_{j}+\frac{kT_{j}}{m}\frac{\partial 
P^{N}}{\partial {\bf U}_{j}})\,+\,\gamma_{ij}P^{N}\frac{1}{T_{j}}\frac{\partial 
T}{\partial {\bf R}_{j}}].
\end{equation}

This is the N particle Fokker Planck equation we set out to derive from the 
scheme of nonequilibrium thermodynamics as applied to systems with internal 
degrees of freedom. In the presence of a temperature gradient a coupling occurs, 
as could be expected, to the value of that gradient at the position of each 
particle.

In our derivation of Eq. (\ref{eq:e19}), direct interactions between the 
Brownian particles were completely neglected. It must be pointed out, however, 
that this derivation can easily be extended to the case that direct interactions 
$\varphi_{ij}\,=\,\varphi(\vert {\bf R}_{i}-{\bf R}_{j}\vert)$ play a 
significant role. 

To perform that extension the local equilibrium disatribution function 
(\ref{eq:c4}) must be adequately modified replacing Eq. (\ref{eq:c2}) by

\begin{equation}\label{eq:e20}
H_{i}\,=\,\frac{1}{2}mU_{i}^{2}\,+\,\frac{1}{2}\sum_{j}\varphi_{ij}\; ,
\end{equation}

\noindent thus  attributing one half of the interaction energy of particle i and 
particle j to particle i and the other half to particle j. It is then not 
possible anymore to determine $f_{i}$ explicitely in a simple manner. However 
with $f_i = \mu_B - p_B\rho^{-1}_e$ and $P_{l.e.}^{N}$ therefore given by

\begin{equation}\label{eq:e21}
P_{l.e.}^{N}\,=\,\prod_{i}N^{-1}\exp\{[m\mu_{B}({\bf 
R}_{i})\,-\,\frac{1}{2}m{\bf 
U}_{i}^{2}\,-\,\frac{1}{2}\sum_{j}\varphi_{ij}\,-\,p_{B}({\bf 
R}_{i})\rho_{e}^{-1}]/kT({\bf R}_{i})\,+\,1\}\; ,
\end{equation}                                                           

\noindent (which reduces, whithout direct interactions and the consequent 
perfect pressure value for $p_{B}$, to Eq. (16)) the analysis can proceed as 
before. One must then take into account the continuity equation obeyed by 
$P^{N}$ in the presence of direct interactions and obtains in a manner 
completely analogous to the one described above (thereby introducing also a new 
modified heat current (cf. Eq. (\ref{eq:d7})) the full Fokker-Plank equation,

\begin{equation}\label{eq:e22}
\frac{\partial P^{N}}{\partial t}+\sum_{i}{\bf U}_{i}\cdot\frac{\partial 
P^{N}}{\partial {\bf R}_{i}}-\sum_{ij}m^{-1}\frac{\partial \varphi_{ij}} 
{\partial {\bf R}_{j}}\cdot\frac{\partial P^{N}}{\partial {\bf 
U}_{i}}\,=\,\sum_{ij}\frac{\partial}{\partial {\bf 
U}_{j}}\cdot[\beta_{ij}(P^{N}{\bf U}_{j} + \frac{kT_{j}}{m}\frac{\partial 
P^{N}}{\partial {\bf U}_{j}})+\gamma_{ij}P^{N}\frac{1}{T_{j}}\frac{\partial 
T}{\partial {\bf R}_{j}}].
\end{equation}

The Fokker-Plank equation (\ref{eq:e22}) now remains applicable to systems which 
are not necessarily dilute.

\section{Conclusion and discussion}

In the preceeding sections we considered an ensemble of Brownian particles 
suspended in a nonuniform heat bath as a system with internal degrees of freedom 
and applied to it the scheme of nonequilibrium thermodynamics.We showed that it 
is then a simple matter to obtain the Fokker Planck equation for simultaneous 
Brownian motion of N particles in the presence of a temperature gradient. In 
fact the derivation along these lines turns out to be no more involved than the 
corresponding one for a single Brownian particle, even though many particle 
hydrodynamic interactions have to be taken into account. It is the simplicity of 
the nonequilibrium thermodynamic approach which forms its main attraction, in 
particular when it concerns, as it does in this case, phenomena which occur 
somewhere on the interface between macroscopic and microscopic physics.

This thermodynamic approach should be compared to statistical mechanical methods 
to derive the basic equation describing many particle Brownian motion. In an 
equilibrium heat bath statistical theories where given by Deutch and Oppenheim 
\cite{kn:deutch} and Murphy and Aguirre \cite{kn:murphy}. While the statistical 
mechanical treatments are more satisfactory from a fundamental point of view, 
they do not seem in the present stage to provide substantially more detailed 
results. 

It should be noticed that in the application given above the extension of 
nonequilibrium thermodynamics to systems with internal degrees of freedom 
already contains elements of statistical physics beyond those present when the 
concept was introduced \cite{kn:prigogine}. Thus the theory developed here makes 
use of the probability density function in the phase space of the N Brownian 
particles and of Gibbs entropy postulate to relate the entropy of the system to 
this distribution function.

In their original paper Prigogine and Mazur stated that it was their object, in 
considering irreversible processes connected to internal degrees of freedom, to 
extend the domain of applicability of nonequilibrium thermodynamics 
(d'\'{e}tendre le domaine d'application de la thermodynamique des phenom$\grave{e}$nes 
irr\'{e}versibles). The question may then be put forward if and why such an 
extension gives rise to a correct description of irreversible phenomena, as it 
obviously does in the case studied in this paper. The answer is that an 
extension to internal degrees of freedom is legitimate as long as the 
irreversible processes connected to these variables are slow in comparison to a 
molecular time scale. Think however for a moment that the N particles considered 
are neither large nor heavy compared to bath molecules but form a marked dilute 
set of these. In that case the Fokker-Planck equation derived makes no sense for 
a description of the motion of the molecules concerned, which is such that the 
Maxwell velocity distribution is attained on a molecular time scale. However, 
the (slow) long time diffusion regime of the relevant particles may still be 
described by standard nonequilibrium thermodynamics. We have an illustration 
here of the limitations for extending that theory: it is an illusion that a 
small number of added pseudo-thermodynamic variables, whatever their nature , 
may suffice to adequately describe (with a fundamentally macroscopic discipline) 
phenomena taking place on the time scale of molecular, or microscopic, events. 
However when one is confronted with a separation of time scales and some 
phenomena take place at intermediate times, a proper extended thermodynamic 
nonequilibrium theory may demonstrate its value in providing the means to easily 
describe and categorize, together with their symmetries, the irreversibilities 
which take place. 

\section{Acknowledgement}
P.M. acknowledges a useful discussion whith Dick Bedeaux. This work has been 
partially supported by DGICYT of the Spanish Government under Grant No. PB 
95-0881.


\newpage

\begin{thebibliography}{99}

\bibitem{kn:agusti} A. P\'erez-Madrid, J. M. Rub\'{\i}, and P. Mazur, Physica A 
{\bf 212}, 231 (1994).

\bibitem{kn:groot}S. R. de Groot and P. Mazur,Non-Equilibrium Thermodynamics 
(Dover, New York, 1984).

\bibitem{kn:zubarev}D.N. Zubarev and A. G. Bashkirov, Physica {\bf 39}, 334 
(1968).

\bibitem{kn:shea}J. Shea and I. Oppenheim, J. Phys. Chem. {\bf 100}, 19035 
(1996).

\bibitem{kn:prigogine}I. Prigogine and P. Mazur, Physica {\bf XIX}, 241 (1953).

\bibitem{kn:murphy}T.J. Murphy and J. L. Aguirre, J. Chem. Phys. {\bf 57}, 2098 
(1972).

\bibitem{kn:deutch}J. M. Deutch and I. Oppenheim, J. Chem. Phys. {\bf 54}, 3547 
(1971).

\end{thebibliography}
\end{document}